\pdfoutput=1

\documentclass[aps,prl,twocolumn,groupedaddress,amsmath,amssymb, longbibliography]{revtex4-1}
\usepackage{graphicx}
\usepackage{float} 
\usepackage{xcolor}
\usepackage{soul}

\begin{document}

\title{A Cascade Leading to the Emergence of Small Structures in Vortex Ring Collisions}

\author{Ryan McKeown$^1$, Rodolfo Ostilla-M\'onico$^{1,2}$, Alain Pumir$^3$, Michael P. Brenner$^1$, and  Shmuel M. Rubinstein$^{1,*}$}

\affiliation{$^1$ School of Engineering and Applied Sciences, Harvard University, Cambridge, MA 02138, USA \\
$^2$ Department of Mechanical Engineering, University of Houston, Houston, TX 77204, USA \\
$^3$ Universit\'e de Lyon, ENS de Lyon, Universit\'e Claude Bernard, CNRS, Laboratoire de Physique, F-69342, Lyon, France \\
$^*$ Corresponding Author: shmuel@seas.harvard.edu }

\date{\today}

\begin{abstract}
When vortex rings collide head-on at high enough Reynolds numbers, they ultimately annihilate through a violent interaction which breaks down their cores into a turbulent cloud. We experimentally show that this very strong interaction, which leads to the production of fluid motion at very fine scales, uncovers direct evidence of a novel iterative cascade of instabilities in a bulk fluid. When the coherent vortex cores approach each other, they deform into tent-like structures, and the mutual strain causes them to locally flatten into extremely thin vortex sheets. These sheets then break down into smaller secondary vortex filaments, which themselves rapidly flatten and break down into even smaller tertiary filaments. By performing numerical simulations of the full Navier-Stokes equations, we also resolve one iteration of this instability and highlight the subtle role that viscosity must play in the rupturing of a vortex sheet. The concurrence of this observed iterative cascade of instabilities over various scales with those of recent theoretical predictions could provide a new mechanistic framework in which the evolution of turbulent flows can be examined in real-time as a series of discrete dynamic instabilities.
\end{abstract}


\maketitle

Our work experimentally and numerically revisits a classical study reported in 1992 by Lim and Nickels, wherein they investigated the instability of vortex filaments during the head-on collision of two dyed vortex rings ~\cite{Lim:1992}. They demonstrated that the collision of a red vortex ring with a blue vortex ring gives birth to a tiara of smaller, half-red, half-blue Janus vortex rings, mediated by reconnections of vortex cores. Additionally, Lim and Nickels insightfully remarked that when the the vortex ring collisions are more vigorous, no secondary rings are created; instead, a turbulent cloud forms nearly instantaneously. This turbulent cloud is composed of a multitude of small-scale flow structures, the dynamics of which were too rapid to capture experimentally or numerically at that time. 

Vortex tubes can be observed in many flow configurations, from the largest scales~\cite{Batchelor:1970,Saffman:1992} down to the finest scales in turbulent flows \cite{Douady:1991,Siggia:1981,Jimenez:1993, Ishihara:2009}. Their interactions are mediated by many instabilities and may lead to breakdown, reconnection, and annihilation of vortex lines due to viscosity. Vortices can be considered topologically protected, in the sense that without viscosity, a vortex line cannot be broken ~\cite{Kida:1987,Kleckner:2013,Kerr:2018,McGavin:2018}. Thus, the annihilation of vortex lines is impossible in the infinite Reynolds number limit (or equivalently, in the zero-viscosity limit). At moderate Reynolds numbers, laminar reconnections dominate vortex line interactions ~\cite{Lim:1992,Oshima:1977,Schatzle:1987,Ashurst:1987,Melander:1989,Saffman:1990,Shelley:1993}. However, at higher Reynolds numbers, interactions become more violent and lose their laminar character ~\cite{Lim:1992,Leweke:1998}. The rapid disintegration of coherent vortex structures and formation of small-scale vortices are likely mediated by a combination of various well-studied vortex instabilities ~\cite{Crow:1970,Tsai:1976,Bayly:1986,Waleffe:1990,Kerswell:2002,Laporte:2000,Leweke:2016}. Nevertheless, many questions remain regarding the dynamics that initiate the breakdown of vortices from large to small scales.

 In this work, we examine the conceptually simple configuration of two identical vortex rings colliding head-on~\cite{Oshima:1978}. The early experimental results of~\cite{Lim:1992} indicate---for collisions at sufficiently high Reynolds numbers---the development of a violent interaction between the rings, causing them to rapidly disintegrate into a turbulent cloud. Due to the very small temporal and spatial scales involved in this interaction process, it has not been possible to study in detail the interaction that initiates the complete breakdown of the vortices. By using state-of-the art visualization techniques, we investigate with sufficient temporal and spatial resolution how this violent interaction breaks down the colliding vortex rings.

From a fundamental point of view, understanding the formation of small-scale flow structures over very short times is a classical problem in fluid mechanics~\cite{TG:1937}, as it is expected to play an important role not only in vortex reconnection, but more generally in turbulent flows~\cite{Frisch:1995}. It has been recognized that the close-range interaction of vortices is also a prime candidate for the formation of singularities in the inviscid 3D fluid equations, although the numerical evidence for or against the existence of such solutions has been ambiguous~\cite{Constantin:2007,Brachet:1983,Pumir:1990,Kerr:1993,Hou:2006}.

Recently, a new class of mechanisms for energy transfer has been proposed, in which the kinetic energy of a flow is conveyed from large to small scales via an iterative cascade, with the same elementary process repeating again and again on smaller and smaller scales \cite{Brenner:2016, Tao:2016}, reminiscent of early simulations \cite{Pumir:1987,Pumir:1990,Kerr:1993,Kerr:2013}. Analogous iterative instabilities have been previously observed in the breaking of fluid jets into droplets \cite{Brenner:1994, Shi:1994}. The physical realization of this iterative process leading to the emergence of small flow structures is envisioned through the interaction between two antiparallel vortex filaments \cite{Siggia:1985,PumirSig:1987}, whose collision is described by a universal similarity solution \cite{Brenner:2016}. The filaments become perturbed and develop into a characteristic shape reminiscent of two opposing tents. The collision initiates at the nose of the tents, leading to what will be referred throughout the text as tent-like structures \cite{Brenner:2016}. As a result of the collision, the filaments flatten into extremely thin vortex sheets. An iterative cascade of instabilities occurs if the sheets, themselves, break up into smaller vortex filaments which subsequently collide, flattening and breaking down into even thinner filaments. Estimates show that these dynamics occur over fleeting time-scales and diminutive length scales \cite{Brenner:2016}. To date, it has not been possible to directly observe the extreme flattening of colliding vortex cores, followed by their subsequent breakdown into secondary and tertiary filaments, as documented here.

\begin{figure}[ht] 
\includegraphics[width = 0.5\textwidth]{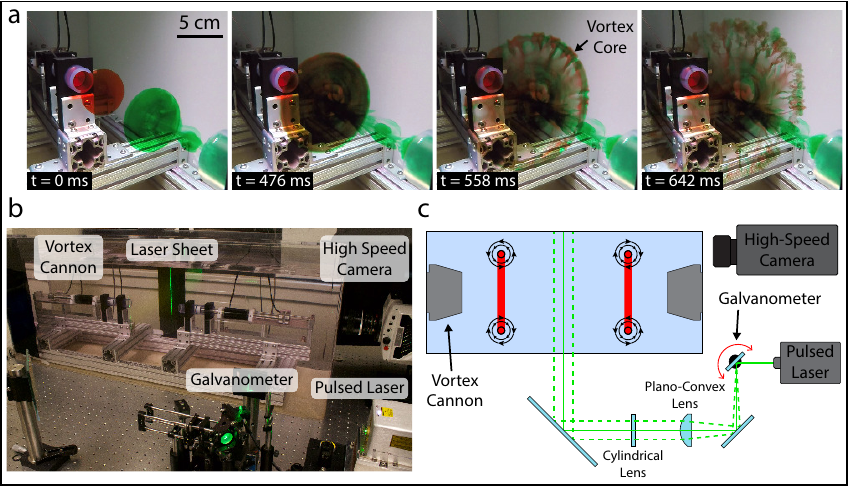}
\caption{Experimental system. (a) A montage of images showing the head-on collision of two vortex rings formed at Re = 8000 and SR = 2.5, where both rings are dyed with food coloring. Upon first colliding, the vortex rings stretch and grow radially before rapidly breaking down to fine-scale ``smoke.'' (b) An image and (c) schematic of the two vortex cannons and high-speed scanning laser sheet fluorescence microscopy setup, which enables each scan of the flow to be reconstructed into a 3D volume.   
\label{Fig1}}
\end{figure} 

Directly observing the breakdown of large coherent vortices into a turbulent cloud requires the visualization of large flow structures disintegrating over ephemeral time-scales  (a few milliseconds) into the smallest constituents of turbulent flow (micrometer length scales). The difficulty results from the chaotic spatiotemporal nature of these processes, making them hard to pin down. In order to isolate and experimentally probe the emergence of the turbulent cloud that results from the breakdown of colliding vortices, we examine the head-on collision of two identical vortex rings, as shown in Fig.~\ref{Fig1}(a) and video 1 \cite{Lim:1992}.  

The planar geometry of the  vortex ring collision restricts all of the dynamics to occur within a narrow volume whose position is fixed in the laboratory frame. This confinement allows for the real-time, high speed, and fully three-dimensional visualization of the flow. Our experimental setup is shown in Fig.~\ref{Fig1}(b-c) and described in detail in Appendix A. 

Two vortex rings are fired head-on using a piston-cylinder assembly in a 75-gallon water aquarium ($45 \times 122 \times 50$ cm$^3$). The vortex cannons are capable of reaching a maximum Reynolds number, $\text{Re}=UD/\nu$, of 25,000 and a maximum stroke ratio, SR = $L/D$, of 4 \cite{Gharib:1998}. $U$ and $L$ are the piston velocity and stroke length, respectively, $\nu$ is the kinematic viscosity of the fluid, and $D = 2.54$ cm is the tube diameter. The cores of the vortex rings, where the vorticity of the flow is concentrated, are dyed \cite{Schatzle:1987} with a fluorescent dye (Rhodamine B) in order to track their motion and deformation. Complementary two-dimensional particle-image velocimetry (PIV) measurements were conducted to ensure that the injected dye coincides with the vortex cores during the collision, as discussed further in Appendix B. The breakdown dynamics of the cores are directly visualized in real time and in full 3D. The collision plane is illuminated by a pulsed ($\approx 15$ ns), $2$-Watt laser sheet (Spectraphysics Explorer One 532-2W), synchronized with the exposure signal of a high-speed imaging sensor (Phantom V2511). The laser sheet scans through the flow over a distance of up to $2.54$ cm at a frequency of $1$ kHz, allowing a data capture rate of $1,000$ volumes per second; this high scanning rate ensures that the dynamics of the flow are effectively ``frozen'' in each individual scan. The full 3D spatiotemporal dynamics of the vortex cores are reconstructed with Dragonfly visualization software (Object Research Systems) at a maximal spatial resolution of $145$ $\mu$m along the collision plane and $100$ $\mu$m in the scanning direction.

\begin{figure}[ht!] 
\includegraphics[width = 0.5\textwidth]{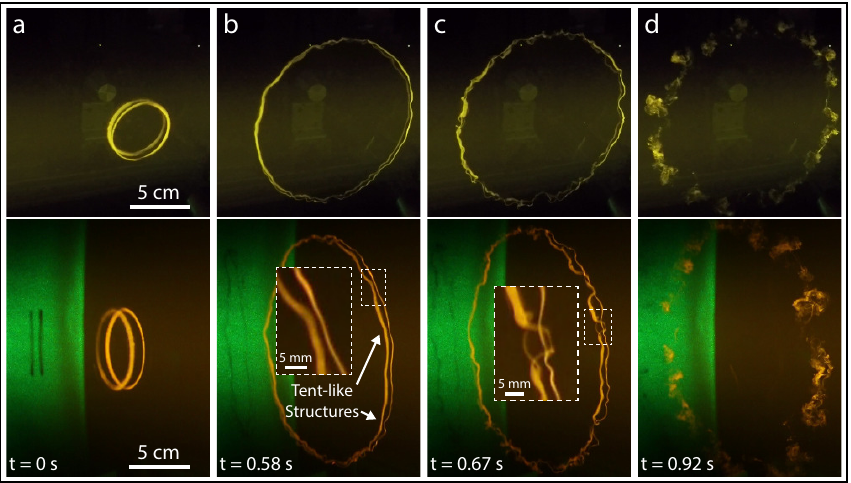}
\caption{Two vortex rings colliding. Four consecutive snapshots simultaneously taken from the front (top) and side (bottom) showing the progressive stages of the head-on collision between two identical vortex rings with Re = 4000 and SR = 2.5. Both cores are injected with a small volume of fluorescent dye (Rhodamine B) and illuminated from the side by a scanning green laser sheet, highlighting the core in bright yellow. (a) As the two cores approach, they stretch radially but remain circular. (b) With further stretching, the vortex rings develop long-wavelength perturbations that form tent-like structures. (Inset) Zoomed-in view of a developing tent-like structure. (c) The tips of the tent-like structures flatten due to the intense strains exerted by the circulating cores. (Inset) The flattened core stretches into a vortex sheet and splits into two secondary vortex filaments. (d) The initial breakdown of the vortex cores at these local tent-like structures propagate along the vortex cores, leading to the annihilation of the vortex rings and the formation of a turbulent cloud.      
\label{Fig2}}
\end{figure} 

The collision between two vortex rings occurs over several distinct stages, shown for a typical example in Fig.~\ref{Fig2} and video 2. As the two rings initially approach one another, they expand radially, stretching along the collision plane, as shown in Fig.~\ref{Fig2}(a). This stretching is initially uniform, such that the rings maintain a toroidal shape. Eventually, however, azimuthal undulations develop circumferentially into unstable tent-like structures around the vortex rings, as shown in  Fig.~\ref{Fig2}(b-c). The undulations grow, contact, and initiate the complete breakdown of the coherent cores into a tiara of turbulent puffs, shown in Fig.~\ref{Fig2}(d). 

The observed initial azimuthal undulations of the core can arise from two different mechanisms \cite{Leweke:2016}: the Crow instability \cite{Crow:1970} and the elliptical instability \cite{Moore:1975, Tsai:1976}. The Crow instability develops from the interaction of the two vortex rings and has a wavelength on the order of the distance between the rings, much larger than the core radius, $\sigma$. In contrast, the elliptical instability occurs from a resonant interaction of a single vortex ring with a strain field, either from itself or from the other vortex, and has a wavelength on the order of the core size. The current experiment has an initial instability that is consistent with the Crow instability, which leads to the initial formation of tent-like structures. As the Reynolds number increases, we expect that the ensuing dynamics will become richer and more complicated as a result of the interplay between the different types of instabilities \cite{Leweke:2016}. 

\begin{figure*}[ht] 
\includegraphics[width = \textwidth]{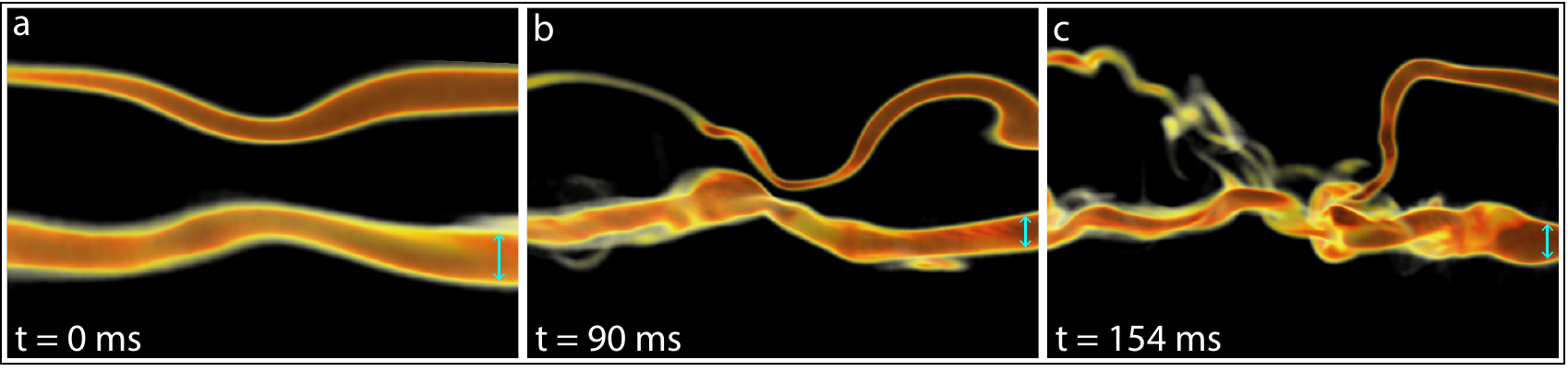}
\caption{3D reconstruction of the breakdown of two dyed vortex cores. Three snapshots showing the late-stage dynamics of two colliding vortex rings where Re = 4000 and SR = 2.5. (a) The cores deflect toward one another, developing a tent-like structure. (b) The tent-like structure grows in amplitude, and the lower core flattens into a vortex sheet at the region where the two cores are closest. (c) The cores contact and break down. In all panels, the cyan arrows indicate length scales: (a) $1.7$ mm, (b) $1.4$ mm, and (c) $1.4$ mm.     
\label{Fig3}}
\end{figure*} 

The nonlinear dynamics---following the initial formation of the long-wavelength undulations---develop the two vortex cores into opposing tent-like structures, as shown by the 3D reconstruction of a typical collision of two dyed filaments in Fig.~\ref{Fig3}(a-c) and in video 3. The tent-like structures grow as a result of the mutual strain imposed by the two interacting vortex cores. At the point of shortest separation, the strain is the strongest, which amplifies the local curvature of the filaments. Eventually, the mutual strain becomes so strong that the lower core flattens into a vortex sheet, as shown in Fig.~\ref{Fig3}(b). This vortex sheet flattens to a thickness that is approximately one tenth that of the initial core diameter, consistent with theoretical predictions \cite{Brenner:2016}. Ultimately, the vortex sheet breaks down completely, as shown in Fig.~\ref{Fig3}(c). 

\begin{figure}[hb!] 
\includegraphics[width = 0.5\textwidth]{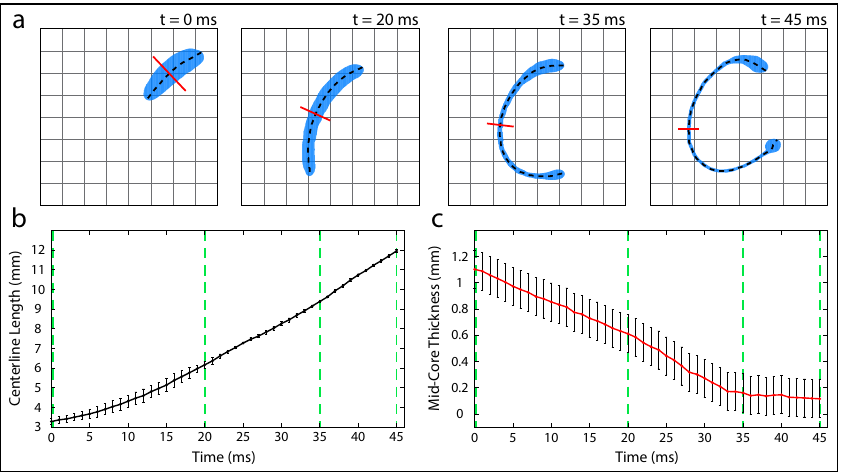}
\caption{ Extreme stretching of a colliding vortex core. (a) Cross-sectional slice of a deforming vortex core extracted from 3D flow visualization during the collision of a dyed vortex ring with an invisible, undyed vortex ring. The dyed vortex core, rotating in the clockwise direction, is shown in blue and is stretched by the undyed vortex core, which rotates in the counter-clockwise direction. The centerline of the vortex core is indicated by the black dashed line, and the plane perpendicular to the midpoint of the centerline is shown by the red lines. The grid spacing is 1 mm $\times$ 1 mm. (b) Centerline length vs. time for the dyed core. (c) Core thickness along the midpoint of the centerline vs. time. The green dashed lines correspond to the times indicated in (a), the data in both plots is averaged over 10 adjoining slices, and for this collision, Re = 4000 and SR = 2.5.
\label{Fig4}}
\end{figure}

\begin{figure*}[p] 
\includegraphics[width = \textwidth]{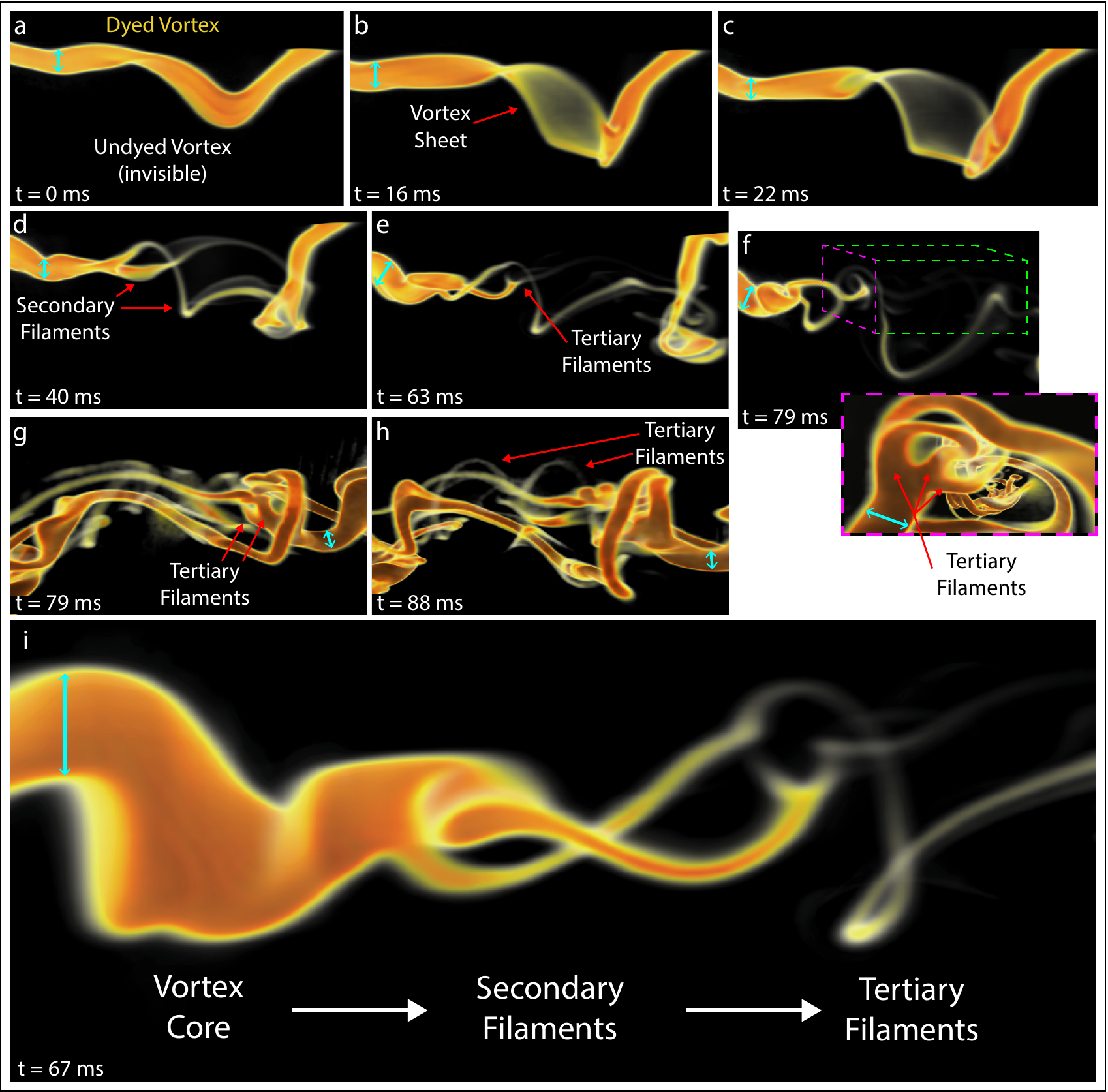}
\caption{An iterative cascade of instabilities leads to the emergence of small-scale flow structures. 3D reconstruction of the breakdown dynamics following the collision of a dyed vortex ring (top) with an invisible undyed one (bottom). For this typical example, Re = 4000 and SR = 2.5. (a) The upper dyed vortex develops a tent-like perturbation that deflects toward the undyed lower vortex. (b) The tent-like structure flattens into a very thin vortex sheet. Most of the dye collects at the edges of the sheet. (c) The thin vortex sheet ruptures, and a hole is formed, which extends down the rest of the core. (d) The vortex sheet splits into two smaller secondary filaments, which unravel the vortex core at both ends. (e) A secondary vortex splits into smaller tertiary vortex filaments. (f) A complex structure of interacting secondary and tertiary vortex filaments emerges. (Inset) Zoomed-in and contrast-enhanced view showing a network of interacting coherent secondary and tertiary vortex filaments. The magnified view is a sub-volume of the main panel indicated by the dashed box and is viewed through the dashed magenta window. (g-h) Full view of the emerging tertiary filaments. The contrast is enhanced to highlight the faint tertiary filaments. (i) The essence of the dynamics is captured entirely in one snapshot, simultaneously showing three generations of the iterative cascade. In every panel, the size of a distinct feature is indicated by a cyan scale bar: (a) $1.3$ mm, (b) $1.2$ mm, (c) $1.2$ mm, (d) $1.7$ mm, (e) $2.9$ mm, (f) $2.1$ mm, (Inset) $1.1$ mm, (g) $1.1$ mm, (h) $0.6$ mm, and (i) $2.8$ mm.\label{Fig5}}
\end{figure*}

Imaging the dynamics beyond the formation of tent-like structures requires increased spatial and temporal resolution. The dynamics of both vortex cores are rapid and three-dimensional, making them difficult to distinguish clearly when both cores are dyed. However, they are clearly discernible when only one cores is injected with dye while the other remains invisible, as shown for a typical example in Figs.~\ref{Fig4}-\ref{Fig5}. 

The initial breakdown of the dyed vortex core can be characterized by slicing through the 3D reconstruction and examining the deformation of the core's cross section, as shown in Fig.~\ref{Fig4}(a). The vortex core stretches and deforms into a curved vortex sheet as a result of the strain exerted upon it by the counter-rotating undyed core, as shown in Fig.~\ref{Fig4}(a). The centerline length of the vortex core grows linearly to nearly four times its initial length, as shown in Fig.~\ref{Fig4}(b). While the stretching of the dyed core is initially uniform, when the end-to-end distance of the core reaches three times its initial length, the ends of the core bulge as the dye and likely the vorticity are continuously drawn from the center of the sheet to the edges. Concomitantly, the center of the vortex sheet contracts and thins until its thickness reaches our spatial resolution limit, as shown in Fig.~\ref{Fig4}(c). As a result of this intense stretching and thinning of the vortex sheet, the aspect ratio of the core approaches a value of 100:1 before the dye in the center of the vortex sheet becomes too dilute to resolve and the bulges at the edges of the sheet roll up into secondary vortex filaments.  

During the later stages of the breakdown, however, the colliding vortex cores undergo far more complicated three-dimensional dynamics, as shown in Fig.~\ref{Fig5} and videos 4-5. Initially, the dyed vortex is distorted by the interaction with the partner vortex and forms a tent-like structure, as shown at $t=0$ ms in Fig.~\ref{Fig5}(a). This interaction causes the vortex to flatten into an extremely thin vortex sheet, as shown in Fig.~\ref{Fig5}(b). The vortex sheet continues to thin, and the dye gradually collects at the edges, as shown in Fig.~\ref{Fig5}(c). Eventually, the intensity of the vorticity in the center the sheet becomes very small, and concentrates into two new filaments emerging from the edges of the sheet, as shown in Fig.~\ref{Fig5}(d). The secondary filaments undergo complicated three-dimensional motion as they unravel the original vortex core. Eventually, the daughter filaments, themselves, split into even thinner tertiary filaments, as indicated in Fig.~\ref{Fig5}(e-f). The tertiary filaments are so thin, that they are difficult to resolve, but enhancing the contrast reveals a complex topology of fine, multi-scale filaments, as shown in the inset of Fig.~\ref{Fig5}(f). The tertiary filaments continue to stretch and interact with each other, the secondary filaments, and with the undyed vortex, as shown in Fig.~\ref{Fig5}(g-h).

The essence of the overall breakdown mechanism is captured elegantly in a single moment which simultaneously showcases three generations of vortices in the iterative cascade, as shown in Fig.~\ref{Fig5}(i)---the primary vortex core splits into secondary filaments, which then split again into tertiary filaments. The fluorescent dye that is injected into the vortex cores effectively traces their position and slightly under-estimates their size, defined by the vorticity distribution in the cores detailed in Appendix B. The mean thickness of the initial primary, secondary, and tertiary cores is approximately $1.83\pm0.1$ mm, $0.61\pm 0.1$ mm, and $0.22\pm 0.1$ mm, respectively. The same intensity threshold is maintained during the measurement of each core size, and the topology of the filaments is quite stable over a wide range of intensity thresholds. Thus, at each iteration of the breakdown, the vortex core size decreases by a factor of three. As the iterative breakdown process generates smaller and smaller scales of vortices, it is ultimately halted by viscosity. This therefore suggests that the iterative cascade could continue and repeat itself to even smaller scales during higher Reynolds number collisions.

Nevertheless, it is important to point out that viscous effects play a critical role even in the breakdown of the primary vortex core. Viscous effects are required for the topological change in which the vortex sheet splits into secondary filaments, clearly shown in Fig.~\ref{Fig5}(b-c). Viscosity acts only for short times and on small scales when the sheet is stretched thinly and appears to rupture. Additionally, it is important to stress that three-dimensional effects play an essential role in the breakdown dynamics. While the splitting of the vortex sheet into smaller filaments may seem to be a quasi-two-dimensional phenomenon, the relative motion of the vortex filaments is induced both from individual vortex filaments, from interactions with neighboring filaments, and vortex stretching.

\begin{figure}[hbt!] 
\includegraphics[width = 0.5\textwidth]{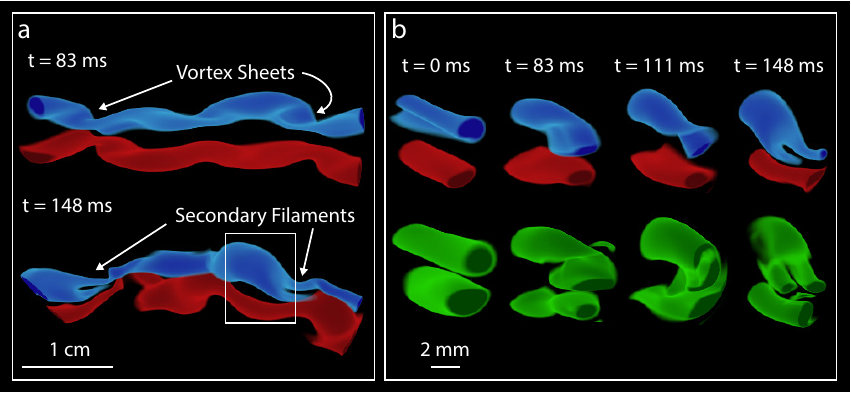}
\caption{Numerical simulation of colliding vortex rings. (a) Volumetric plot of dye concentration in a simulation of two colliding vortex rings whose cores are initially seeded with red and blue dye, respectively. As the vortex cores approach, the blue vortex flattens into thin vortex sheets, which break down into secondary filaments. (b) Zoomed-in, cross-sectional view of the developing secondary filaments indicated by the white box in (a). The top panels plot the dye concentration, and the bottom panels plot the vorticity modulus. The formation and rupture of sheets into filaments is clearly shown by the top (blue) vortex, while the lower vortex shows that if no rupture occurs, the sheet eventually curls back into a vortex when the strain is released.     
\label{Fig6}}
\end{figure} 

To probe this breakdown mechanism further, we also examine the head-on collision of two vortex rings with direct numerical simulations of the Navier-Stokes equations. Technical details of the simulations are described in Appendix C. In the same configuration as the experiments, the initial condition for the simulations consists of two identical vortex rings colliding head-on. In the calculation shown here, the Reynolds number, defined by $\text{Re}_{\Gamma} = \Gamma/ \nu$, where $\Gamma$ is the circulation of the rings, is $\text{Re}_{\Gamma} = 3500$. The initial core slenderness ratio, $\Lambda = \sigma/R$, is set at 0.35, where $R$ is the vortex ring radius. Throughout the progression of the simulation, we calculate both the evolution of the vorticity distribution as well as the concentration of an advected dye in both vortex rings. Computational limitations prohibit using a dye diffusivity comparable to that of the experiments, $\approx \, 10^{-3} \, \nu$ \cite{Gendron:2008}. The simulated dye diffusivity is equal to the kinematic viscosity of the fluid. The initial dynamics mirror those of the experiments, as shown in Fig.~\ref{Fig6}(a)---the rings expand radially as they approach one another and develop a long-wavelength Crow instability \cite{Leweke:2016}. 

The simulations nicely capture one iteration of the cascade mechanism, as shown in Fig.~\ref{Fig6}(b) and video 6. The dye tracks the vorticity during the initial development of the instability and formation of the tent-like structures. The dynamics of the simulations are qualitatively similar to those of the experiments as one of the dyed vortex cores flattens into a sheet, which then ruptures into two secondary filaments. These same dynamics are mirrored in the vorticity distribution, as shown in green in Fig.~\ref{Fig6}(b). The circulation is distributed between the two ruptured filaments in a 2:1 proportion, and its sum roughly corresponds to the pre-split sheet configuration, with minor losses ($<1\%$), likely due to viscosity. The diffusive losses for the dye are larger: $\approx 40\%$ (a similar phenomena is also seen by the evanescent secondary filaments in Fig.~\ref{Fig5}(i)), but the ratio between the amount of dye in both secondary filaments matches those of the circulations. The simulations suggest that the minimum thickness of the sheet joining the two secondary filaments is given by the viscosity of the fluid. If $\dot{\gamma}$ is the shear rate through which the sheet is being stretched, viscous effects will set in when the sheet thickness is of order $\sqrt{\nu/\dot{\gamma}}$. This viscous length scale is calculated to be approximately 0.005R for the vortex sheet at $t=111$ ms, the same order as the thickness of the sheet $(\approx 0.01R)$. This is approximately 4 grid points. While the rupture of the sheet is not produced in a strict sense---i.e. no region with zero vorticity appears---at later times, the vorticity in the sheet remnants becomes vanishingly small and the flow structure can be effectively thought of as two discrete secondary filaments. Vorticity amplification is primarily localized to the breakdown of the vortex cores, while the mean vorticity of the simulation is amplified only slightly. Additionally, the whole collision process tends to make the vorticity lose its initially preferential orientation in the azimuthal direction, indicating the emergence of a turbulent cloud like in the experiments. In Appendix D, we show more details about the evolution of vorticity in the simulations.

Vortex ring collisions at high Reynolds numbers lead to a near-instantaneous breakdown of the initial vortices into a turbulent cloud. We have shown that the formation of this cloud is initiated by an iterative cascade of instabilities: colliding vortex filaments break down into smaller secondary filaments, which then interact with each other and break down into even smaller tertiary filaments. To our knowledge, this is the first reported experimental observation of a dynamic cascade of instabilities in a bulk fluid. The iterative mechanism, whereby large vortex filaments flatten into sheets and then break down into smaller filaments, qualitatively agrees with recent theoretical predictions \cite{Brenner:2016, Tao:2016}, although the mechanism observed experimentally and numerically differs in details from the one envisioned in \cite{Brenner:2016}. The degree to which the vortices stretch during each step of the cascade is a crucial question to understand the possible relevance of the scenario discovered here to the singularity problem in the inviscid equations of fluid motion \cite{Constantin:2007,Fefferman:2006,Moffatt:1994}. A semi-quantitative modelling approach may provide much-needed insight. Viscosity must play an essential, though limited, role at each iteration, as it enables the topological transition that leads to the rupture of the vortex sheets. The observation that vortex ring collisions lead to such complicated dynamics on ever smaller length scales, with viscosity playing such a critical role, therefore exposes the subtlety of understanding the singular or near-singular dynamics of the Euler and Navier-Stokes equations.

Because the breakdown dynamics that emerge in the head-on collision of vortex rings occur locally through the close-range interactions of vortex filaments, one may speculate that it could be extended to other high-Reynolds number flows. In fact, this mechanism could conceivably  supply an effective means for a flow to rapidly convey energy down to the smallest constitutive scales, indicative of the type of dynamics that could lead to the turbulent cascade itself---reminiscent of, but perhaps different from Richardson's initial proposal \cite{Richardson:1922}. While the mechanism uncovered here may be involved in the proliferation of small-scale vortex structures in turbulent flows---documented many times both experimentally \cite{Douady:1991} and numerically \cite{Siggia:1981,Jimenez:1993, Ishihara:2009}---our measurements occur at comparatively moderate Reynolds numbers. We  expect that richer breakdown dynamics have yet to be observed for collisions at higher Reynolds numbers. Nevertheless, our model system provides an exciting new lens through which we can attempt to observe and characterize the emergence of complex, multi-scale flows in real-time. By identifying the mechanisms by which these vortical flows break down to small scales, we hope to develop a new framework for viewing the turbulent cascade as a collection of discrete dynamic instabilities. Our study therefore indicates that the rigorous investigation of colliding vortex rings, at high spatial and temporal resolution, may potentially provide profound insights on both the fundamental physics of vortex breakdown and the deep underlying mathematical foundations that govern the emergence of small scales in violent turbulent flows.

\begin{section}{Acknowledgements}
This research was funded by the National Science Foundation through the Harvard Materials Research Science and Engineering Center DMR-1420570, and through Division of Mathematical Sciences DMS-1411694 and DMS-1715477. M.P.B. is an investigator of the Simons Foundation. S.M.R. acknowledge support from the Alfred P. Sloan Foundation. We dedicate this work to the memory of Leo Kadanoff, who asked whether the turbulent cascade was composed of iterative instabilities more than 20 years ago.
\end{section}

\appendix
\begin{section}{Appendix A: Experimental Setup and Procedure}
Two identical vortex rings are launched head-on into one another in a 75-gallon water aquarium ($45 \times 122 \times 50$ cm$^3$). The vortex rings are formed using two piston-cylinder assemblies, as shown in Fig.~\ref{Fig7}. Each piston is driven by an underwater linear shaft motor (Nippon Pulse S160D) through a stainless steel tube with a diameter, $D = 2.54$ cm, expelling a slug of fluid of length, $L$, through the tapered nozzle. The pistons are impulsively accelerated to a constant speed, $U \leq 1 $ m/s. The displacement of both pistons is measured with a non-contact linear encoder (Renishaw LM15) with a resolution of $5~ \mu$m. The linear motors are servo controlled and coupled with each other through CNC software (TwinCAT3) so that the firing of the vortex rings is synchronized. Each vortex cannon is capable generating vortex rings with a maximum Reynolds number, $\text{Re} = UD/\nu$, of $25,000$, where $\nu$ is the kinematic viscosity of water, and a maximum stroke ratio, SR = $L/D$, of 4 \cite{Gharib:1998}. The outer orifice of the vortex cannon is double-walled, and fluorescent dye (Rhodamine B) is ejected through the thin, $~150 ~\mu$m, gap just prior to firing the pistons, as shown in Fig.~\ref{Fig7} (b) \cite{Schatzle:1987}. As the ejected fluid separates at the sharp edge of the nozzle, the ring of dye becomes trapped in the core of the vortex, allowing us to visualize the dynamics of the vortex cores.  
\begin{figure}[ht] 
\includegraphics[width = 0.5\textwidth]{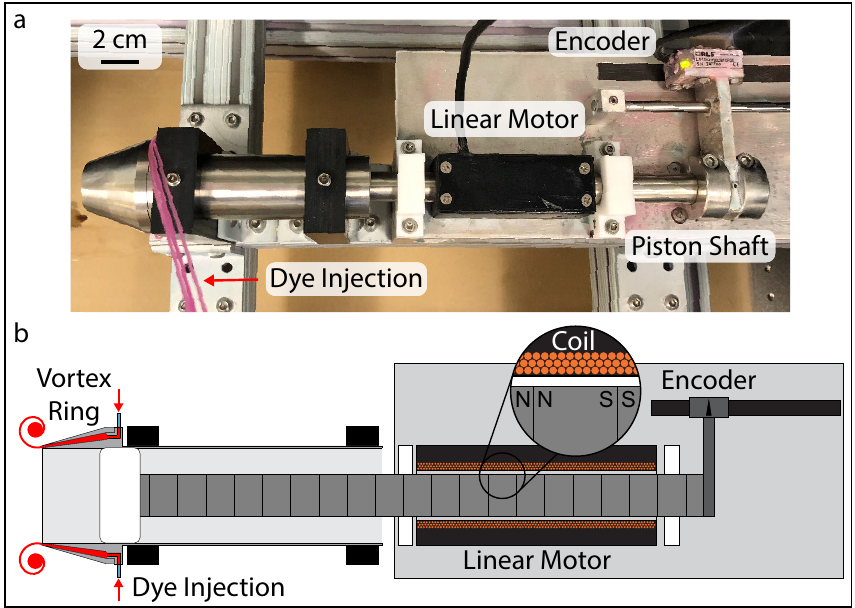}
\caption{Vortex cannon assembly. (a) Image and (b) cross-sectional schematic of the vortex cannon assembly. The piston shaft is loaded with permanent magnets and is impulsively driven by the magnetic fields generated by a series of current-carrying coils in the linear motor. The linear motor is servo controlled, and the feedback position of the piston is measured by a non-contact linear encoder. The core of the vortex ring is dyed by injecting fluorescent dye at the outer edge of the nozzle.   
\label{Fig7}}
\end{figure} 

In all the experiments discussed in this work, the vortex cannons are spaced a distance, $H=8D$, apart. We simultaneously fire two identical vortex rings head-on into each other. As the counter-rotating vortex rings initially approach one another, they expand radially and become flattened along the collision plane, as shown in Fig.~\ref{Fig2} in the main text. We directly observe the breakdown dynamics of the vortex rings in real-time and in 3D by imaging the flow tomographically with a scanning laser sheet. Recently, Irvine et \textit{al}. used a similar visualization technique to measure various properties of coherent vortices in 3D \cite{Kleckner:2013, Scheeler:2017}. They developed a powerful method to recover the flow field of the vortices by mapping the centerlines of the vortex cores, but this required that the cores remain largely undeformed. The dynamics of our system involve extreme contortions and deformations of the vortex cores, which cannot be captured by merely resolving the centerlines of the vortices. Studying this rapid breakdown requires clearly resolving the full three-dimensional structure of the vortex cores.

\begin{figure*}[t!] 
\includegraphics[width = \textwidth]{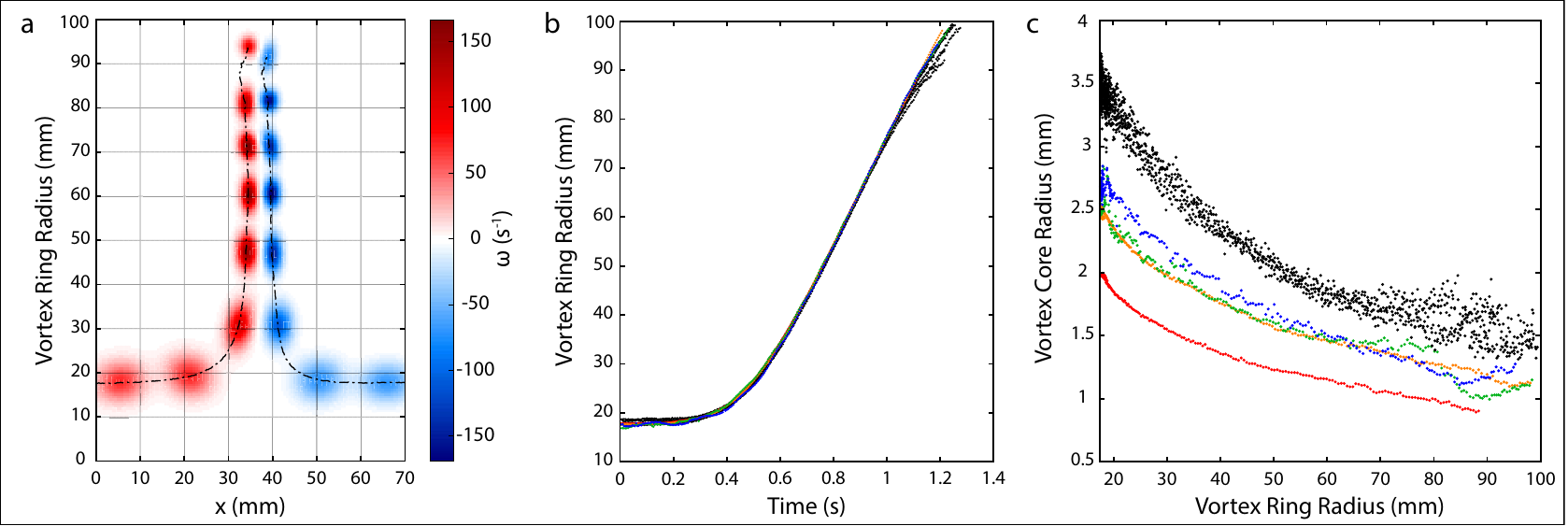}
\caption{Vortex core tracking in 2D with PIV and dye. (a) Vorticity distribution and core trajectory of two colliding vortex rings at several time steps obtained via PIV through a 2D cross section perpendicular to the collision plane. The plotted vorticity distribution is the result of fitting the raw vorticity data to a 2D Gaussian function for each core. As the vortex rings collide, they grow radially and the cores contract, amplifying their vorticity before breaking down. The time steps correspond to t=0, 0.225, 0.5, 0.675, 0.8, 0.9, 1, and 1.25 s, and for all data in this figure Re = 4000 and SR = 2.5. (b) Vortex ring radius vs. time for 10 collisions where the vortex core is tracked from the vorticity data via PIV (black) and from the dyeing the core with fluorescent dye (color). (c) Vortex core radius vs. vortex ring radius for 10 collisions; the core is tracked by fitting to the vorticity data (black) and by injecting the vortex ring with various amounts of fluorescent dye: 0.08 mL (red), 0.1 mL (orange), 0.1 mL (green) and 0.12 mL (blue).
\label{Fig8}}
\end{figure*} 

The collision plane is illuminated by a pulsed ($\approx 15$ ns), $2$-Watt laser sheet (Spectraphysics Explorer One 532-2W), which is synchronized with the exposure signal of a high-speed imaging sensor (Phantom V2511), as shown in Fig.~\ref{Fig1} (b-c) in the main text. The laser beam is deflected by a mirror mounted on a servo-controlled galvanometer (Cambridge Technology Model 6210HM60), which is placed at the focal point of a plano-convex lens; thus, the deflected beam is collimated.  The laser beam then passes through a cylindrical lens that opens into laser sheet with a thickness of $100 ~\mu$m. The laser sheet scans over a distance of up to $2.54$ cm at a frequency of $1$ kHz, driven via a sawtooth command signal. At this rapid scanning rate, the flow is effectively ``frozen'' during each scan. The high-speed camera captures cross sections of the flow illuminated by the laser sheet as it scans through the fluid. A sequential series of 2D image slices is thus continuously captured by the high-speed camera. High-intensity regions in these image slices correspond to the fluorescent dye within the vortex cores. The short pulsing of the laser is critical to prevent the blurring of the images due to the motion of the laser sheet. The dynamics described in this paper cannot be observed with a continuous laser.

The timing of each captured image is correlated with the measured position of the laser sheet. Each stack of 2D image slices is reconstructed into a 3D volume with dimensions of 256 $\times$ 384 $\times$ 114 voxels using Dragonfly visualization software (Object Research Systems). The reconstructed 4D data has a resolution of ($145 \times 145 \times 100$ $\mu$m$^3$ and $1$ msec) in the ($x,y,z$) directions and time. The resolution of the volume in the xy-plane is limited by the magnification of the lens on the high-speed camera (Nikkor f$=85$ mm, f/1.4), and the resolution in the $z$-direction is given by the spacing between the image slices and the thickness of the laser sheet. This imaging technique allows us to directly observe and probe the full volume and resolve the three-dimensional dynamics of the vortex ring collision with high spatial and temporal resolution. 
\end{section}

\begin{section}{Appendix B: Vortex Core Tracking with PIV and Dye}

Complementary PIV measurements were conducted in order to determine how well the fluorescent dye tracks the motion of the vortex cores during the head-on collision. A laser sheet was used to illuminate a 2D cross-section aligned along the central axes of the vortex cannons, perpendicular to the collision plane. Additionally, the aquarium was seeded with polyamide particles with a diameter of 50 $\mu$m and a density of 1.03 g/mL (Dantec Dynamics). Several vortex ring collisions, where Re=4000 and SR=2.5, were imaged with a high speed camera using a window size of 800$\times$1280 at a maximum frame rate of 2000 fps with a resolution of 0.14 mm/pixel. The velocity field along this 2D cross-section was evaluated using the MATLAB program PIVsuite. The locations and sizes of the vortex cores were then detected by fitting the vorticity distribution of each core to a two-dimensional Gaussian function, as shown in Fig.~\ref{Fig8}(a). When the vortex rings collide, they grow radially in an axisymmetric manner, amplifying the vorticity of the core through vortex stretching, before developing perturbations and breaking down, as shown in Fig.~\ref{Fig2} in the main text.

	Additionally, the same collisions were performed and visualized along this illuminated 2D plane with one of the vortex cores dyed using fluorescent dye (Rhodamine B). The cross-section of the dyed core was segmented and fitted to an ellipse in order to extract the centroid of the core, its semi-major axis length, $a$, and its semi-minor axis length, $b$. The trajectories of the colliding vortex cores follow the same linear trend of radial growth for both the cores tracked with vorticity through PIV and through dyeing the core, as shown in Fig.~\ref{Fig8}(b). The vortex core radius was calculated as the average of the standard deviation in the two principal directions for the cores fitted from the vorticity data and the average of $a$ and $b$ for the dyed cores. During the radial growth of the rings when they collide, the vortex cores accordingly contract due to vortex stretching, as shown by Fig.~\ref{Fig8}(c). Naturally, the more dye that is initially injected into the vortex ring, the thicker the core appears throughout the collision. The radius of the dyed core is smaller than that of the core obtained from the vorticity distribution during each stage of the collision. However, for vortex rings injected with at least 0.1 mL of dye, during the later stages of the collision just prior to breakdown---i.e. when the 3D visualizations begin---the core radius from the dye under-estimates the core radius from the vorticity by less than 0.4 mm. This demonstrates that fluorescent dye effectively tracks the motion of the vorticity distribution of the core, especially when the visualized breakdown dynamics of the core begin.
\end{section}

\begin{section}{Appendix C: Numerical Simulation Details}

The incompressible Navier-Stokes equations are solved using an energy-conserving second-order centered finite difference scheme in cylindrical coordinates with fractional time-stepping. A third order Runge-Kutta scheme is used for the non-linear terms and a second order Adams-Bashworth scheme is used for the viscous terms \cite{Verzicco:1996}. The solver uses $q_r=r v_r$ as a primitive variable to avoid singularities near the center axis. The time-step was dynamically chosen so that the maximum Courant-Friedrich-Lewy (CFL) condition number was $1.2$. A Gaussian (Lamb-Oseen) velocity profile was used as the initial condition for the velocities inside the ring with a slenderness of $\Lambda = \sigma/R = 0.35$, where $\sigma$ is the core radius and $R$ is the vortex ring radius. The Reynolds number based on the circulation, $\text{Re}_{\Gamma}$, is $\text{Re}_{\Gamma} = \Gamma / \nu =3500$, where $\Gamma$ is the circulation of the vortex ring. Length and time-scales are matched to the experimental values by using the ring diameter and circulation. The first is matched by assuming the ring diameter coincides with the vortex cannon tube diameter, and the second is matched by using the piston stroke length, the piston velocity, and the correlations developed by Gharib \& Shariff \cite{Gharib:1998}. A passive tracer, corresponding to a dye, is also simulated. Due to computational restrictions, the Schmidt number of the dye is limited to unity. 

The rings are initially positioned at a distance $L_z = 2.5R$ away from each other. A rotational symmetry, $n_{sym}$, of order five was forced on the simulation to reduce computational costs. The ring's initial position was perturbed according to the formula $R(\theta) = R_0 \left( 1 + \epsilon \sin \left(k [\theta+\theta_0 ]\right) \right)$, where $\epsilon$ is taken from a normal distribution with variance of $10^{-2}$ for the first two wavelengths, to account for the ring's self-instability, and $10^{-3}$ for the other wavelengths; $\theta_0$ is taken from a uniform distribution.

The cylindrical computational domain was bounded by stress-free walls at a sufficient distance not to affect the dynamics of the flow. In practice, we set the walls at a distance $R$, below and above the initial locations of the vortex rings, and at a distance of $5R$ from the ring axis in the radial direction. Points were clustered near the collision plane in the axial direction, while uniform resolution was used in the other two directions. A total of $384 \times 512
\times 264$ grid points were used in the azimuthal, radial and axial directions, respectively. Resolution adequacy was checked by monitoring that the viscous dissipation and the energy balance are within the acceptable bound of $1\%$ \cite{van_der_Poel:2015}.

\end{section}


\begin{figure}[hb] 
\includegraphics[width = 0.5\textwidth]{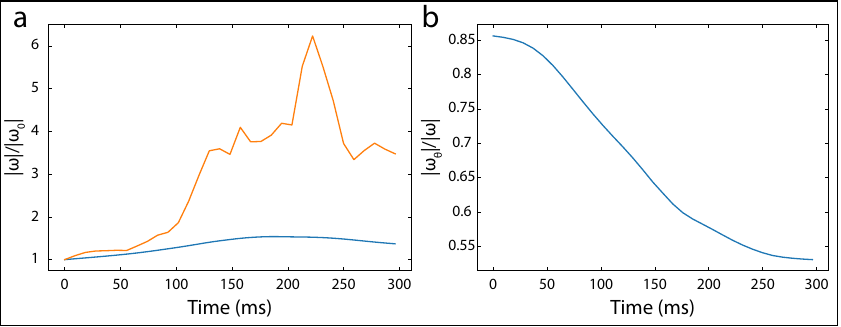}
\caption{Vorticity evolution in the collision simulations. (a) Evolution of the maximum (orange) and mean (blue) vorticity modulus with time, relative to the initial values. (b) Weight of the mean azimuthal component of vorticity relative to the mean vorticity modulus as a function of time. A value of $1/\sqrt{3}\approx 0.577$ would correspond to a vorticity distribution with no preferential direction.
\label{Fig9}}
\end{figure} 

\begin{section}{ Appendix D: Vorticity Evolution in Simulations}

The evolution of the vorticity in the simulated collision was quantified to complement the experimental results. Fig.~\ref{Fig9}(a) shows the amplification of the maximum and mean vorticity modulus during the collision. While the vorticity locally grows by over a factor of five, the mean vorticity never grows beyond 1.5 times its original value. This is because, as shown in Fig.~\ref{Fig5}, not all regions of the ring undergo the iterative evolution shown in Fig.~\ref{Fig6}(b).

Additionally, Fig.~\ref{Fig9}(b) shows how the initially coherent vortex cores, whose mean vorticity is primarily aligned in the azimuthal, or $\hat{\theta}$ direction, become more three-dimensional during the late stages of the collision, as suggested by the experimental observations. Indeed, an isotropic vorticity distribution corresponds to $|\omega_\theta|/|\omega|$ of $1/\sqrt{3}\approx 0.577$, very close to the value attained at the end of the simulation. This demonstrates that as time progresses, the flow becomes fully three dimensional and quasi-isotropic. 

\end{section}

	
  
%

\clearpage

\end{document}